\documentclass[twocolumn,
preprintnumbers,citeautoscript,prl,superscriptaddress,amsmath,amssymb]{revtex4}

\usepackage{hyperref}
\usepackage{graphicx}
\usepackage{times}

\def\SU{\mathrm{SU}}
\def\UU{\mathrm{U}}
\def\SO{\mathrm{SO}}

\begin{document}

\title{Three-dimensional gauge theories with supersymmetry enhancement 
}

\author{Dongmin Gang}
\affiliation{Center for Theoretical Physics, Seoul National University, Seoul 08826, Korea}

\author{Masahito Yamazaki}
\affiliation{Kavli Institute for the Physics and Mathematics of the Universe (WPI), University of Tokyo, Chiba 277-8583, Japan}

\date{\today}

\begin{abstract}
We conjecture infrared emergent $\mathcal{N}=4$ supersymmetry for a class of three-dimensional $\mathcal{N}=2$ $\UU(1)$ gauge theories coupled with a single chiral multiplet. One  example is the case where $\UU(1)$ gauge group has  the Chern-Simons level $-\frac{3}2$ and the chiral multiplet has gauge charge $+1$. Other examples are related to this example either by known dualities or rescaling the Abelian gauge field. We give three independent evidences for the conjecture: 1) exact match between the central charges of the $\UU(1)$ R-symmetry current and the $\UU(1)$ topological symmetry current, 2) semi-classical construction of the $\mathcal{N}=4$ stress-tensor multiplet, and 3) an IR duality  between a direct product of the two copies of the 3d theory on the one hand, and an $\mathcal{N}=4$ theory obtained by gauging the diagonal $\SU(2)$ flavor symmetry of the $T[\SU(2)]$ theory, on the other. The duality in 3) follows from geometrical aspects of the 3d--3d correspondence.
\end{abstract}

\pacs{}
\maketitle


\noindent
{\bf Introduction}

Symmetry has long been a fundamental guiding principle in theoretical physics. One of the most successful examples for this 
is the celebrated supersymmetry \cite{Gervais:1971ji,Golfand:1971iw,Wess:1974tw}, a spacetime symmetry exchanging bosons and fermions. 

Supersymmetry is traditionally regarded as a high-energy symmetry in the ultra-violet (UV). 
There is a different attractive possibility where supersymmetry is emergent in the infra-red (IR)---one starts with a
theory with no supersymmetry in the UV, which flows to an IR fixed point with emergent supersymmetry. Such a possibility 
has actively been studied recently in condensed matter literature \cite{Lee:2006if,Yu:2010zv,Ponte:2012ru,Grover:2013rc,Zerf:2016fti},
and could even be realized experimentally.

One of the virtues of supersymmetry is that it places stringent constraints on the possible physics.
Supersymmetry, however, in itself is not enough for analyzing and better understanding
renormalization group (RG) flow with emergent supersymmetry.
This is because supersymmetry emerges only in the IR, and is not present in the UV theory
which is the starting point of the analysis.
The situation is better if we start with a supersymmetric theory in the UV, and if the theory has emergent {\it supersymmetry enhancement} in the IR:
we then can use the powerful tools from supersymmetry to study the emergence of supersymmetry in itself.

In this Letter we propose examples of such supersymmetry enhancement along the RG flow,
where a class of theories in three space-time dimensions with manifest $\mathcal{N}=2$ extended supersymmetry
has enhanced $\mathcal{N}=4$ supersymmetry in the IR. 

The theory we discuss is rather simple: an abelian Chern-Simons (CS) matter theory coupled to a single chiral multiplet.
In fact, as we will discuss below there are some indications that 
our examples could be ``minimal'' such examples.
That such a simple theory admits supersymmetry enhancement is a surprise, and we hope that 
a through understanding of these examples will provide valuable insights into the emergence/enhancement of supersymmetry in general.
We will below provide three independent evidences for this proposal, 
by taking advantage of several cutting-edge techniques and results for three-dimensional $\mathcal{N}=2$ supersymmetry,
including supersymmetric localization and the 3d--3d correspondence.

\medskip
\noindent{\bf Proposal for $\mathcal{N}=4$ supersymmetry enhancement}

Let us consider a 3d $\mathcal{N}=2$ abelian CS matter theory coupled with a single $\mathcal{N}=2$ chiral multiplet
\begin{align}
\begin{split}
&\mathcal{T}_{k,Q}:=(\textrm{a $\UU(1)$ vector multiplet  with the CS level $k$ coupled} 
\\
& \qquad \qquad \textrm{with a chiral multiplet $\Phi$ of charge $Q$}).
\end{split}
\end{align}
For consistency of the theory \cite{Niemi:1983rq,Redlich:1983kn,Redlich:1983dv} we assume the quantization condition of the bare CS level $k$:
\begin{align}
k \in \mathbb{Z}+ \frac{Q^2}{2}.
\end{align}

The main result of this Letter is to 
propose the following supersymmetry enhancement in the IR:
\begin{align}
\boxed{
\mathcal{T}_{k=-\frac{3}{2}, Q=1} \textrm{ has emergent $\mathcal{N}=4$ supersymmetry in the IR}.} \label{SUSY enhancement}
\end{align}
In the following we shall substantiate this claim by providing three evidences.

\medskip\noindent
{\bf Properties of the IR SCFT}

Let us begin by summarizing some properties of the theory $\mathcal{T}_{k=-3/2,Q=1}$ at the IR fixed point, 
which requires only $\mathcal{N}=2$ supersymmetry manifest in the UV Lagrangian.

We assume that there is no additional emergent Abelian flavor symmetry in the IR. Then, the superconformal $\UU(1)$ R-charge at the IR fixed point can be determined by F-maximization \cite{Jafferis:2010un} and the result is
\begin{align}
(\UU(1)_{R}\textrm{ of }\Phi )= \frac{1}3. \label{IR R-charge}
\end{align} 

The IR SCFT does not seem to have any gauge invariant 1/2 BPS  chiral primary operator (CPO). As we will see later in Table~\ref{Table : local operator spectrum}, we can list local operator spectrum of the theory in the semiclassical limit and do not find any CPO. The same will likely be the case at the IR fixed point,
unless we assume the unlikely possibility of an emergent CPO in the deep IR.
The absence of CPO implies the empty vacuum moduli space for the IR SCFT. 

Using the  UV Lagrangian description with manifest $\mathcal{N}=2$ supersymmetry, the stress-energy tensor central charge $C_T$ \footnote{This is defined from the two-point function of the stress-energy tensor \cite{Osborn:1993cr}, see \eqref{JJ} for similar quantities for conserved currents.} can be evaluated exactly \cite{to-appear}
\begin{align}
\begin{split}
	\frac{C_{T} (\mathcal{T}_{k=-\frac{3}{2},Q=1})}{C_{T} (\textrm{a free chiral $\Phi$})}&= \frac{8}{25}\left(8-\frac{5 \sqrt{5+2 \sqrt{5}}}{\pi }\right)  
	\\&\simeq 0.992549 .
\end{split}
\end{align}
Note that the central charge is even smaller than that of a free chiral theory. 

Suppose the $\mathcal{N}=4$ SUSY enhancement really happens, as our conjecture claims.
Since 3d $\mathcal{N}=4$ theories with Lagrangian description have non-trivial vacuum moduli space, it then follows that the IR  $\mathcal{N}=4$ SCFT does not allow any UV Lagrangian description with manifest $\mathcal{N}=4$ supersymmetry. 
The IR SCFT is therefore a strong candidate for the minimal 3d $\mathcal{N}=4$ SCFT which does not allow for UV Lagragian with manifest $\mathcal{N}=4$ supersymmetry,
where minimal here refers to the smallest non-zero values for the central charge $C_T$.   The situation is analogous to  the 4d case \cite{Maruyoshi:2016tqk,Maruyoshi:2016aim}. In 4d $\mathcal{N}=2$, the Argyres-Douglas theory \cite{Argyres:1995jj} turns out to be the minimal SCFT \cite{Liendo:2015ofa} which allow UV Lagrangian descript  ion only with $\mathcal{N}=1$ supersymmetry.
See also recently found examples of  SUSY enhancements in 4d \cite{Agarwal:2016pjo,Agarwal:2017roi,Benvenuti:2017bpg,Agarwal:2018ejn} and 3d \cite{Gaiotto:2018yjh,Benini:2018bhk}.

\medskip\noindent 
{\bf Evidence 1: $C_{J_R} = C_{J_{\rm top}}$}

Let us next come to the first evidence for our conjecture.

The global symmetries of the theory manifest in the UV Lagrangian are the $\mathcal{N}=2$ R-symmetry $\UU(1)_R$
as well as the topological symmetry $\UU(1)_{\rm top}$, where the conserved current for the latter is
given by 
\begin{align}
J^{\mu}_{\rm top} \propto \epsilon^{\mu \nu \rho} F_{\nu \rho} = \epsilon^{\mu \nu \rho} (\partial_\nu A_\rho - \partial_{\rho} A_\nu) ,
\end{align}
with $A_\mu$ being the dynamical gauge field. These global symmetries are expected to be enhanced to
an emergent $\SO(4)_R$ R-symmetry in the IR so that
\begin{align}
\begin{split}
&\mathbf{4}_{\rm vec} \textrm{ of $\SO(4)_R$ has charges $\left\{ (\pm 1,0),(0,\pm 1) \right\}$} 
\\
&\textrm{under $\UU(1)_R \times \UU(1)_{\rm top}$} . \label{embedding u(1)2 so(4)}
\end{split}
\end{align}

The Weyl-group symmetry of the emergent $\SO(4)$ R-symmetry
contains an emergent $\mathbb{Z}_2$ symmetry exchanging the two global symmetries $\UU(1)_{\rm top}$ and $\UU(1)_{R}$.
The conjecture \eqref{SUSY enhancement} therefore predicts equalities between the correlation functions for the conserved current $J^{\mu}_{\rm top}$ and those for $J^{\mu}_R$. We can in particular consider the case of the two-point functions of the currents, which are determined up to overall constants
$C_{J_{\rm top}}$ and $C_{J_R}$, known as the central charges \cite{Osborn:1993cr}:
\begin{align}
\begin{split}
&\langle J_{\rm top}^{\mu} (x)J_{\rm top}^{\nu} (0)\rangle = C_{J_{\rm top}} \frac{I_{\mu\nu}(x)}{|x|^4},  
\\
&\langle J_{R}^{\mu} (x)J_{R}^{\nu} (0)\rangle = C_{J_R} \frac{I_{\mu\nu}(x)}{|x|^4}, 
\\
&I_{\mu\nu}(x):=\delta_{\mu \nu} - 2\frac{x_\mu x_\nu}{x^2} .
\end{split}
\label{JJ}
\end{align}   
The central charges $C_{J_{\rm top}}$ and $C_{J_R}$ can be computed using supersymmetric localization \cite{Closset:2012vg,to-appear}, and we indeed find their match:
\begin{align}
C_{J_R}=  C_{J_{\rm top}} =  \frac{2}{25} \left(8-\frac{5 \sqrt{5+2 \sqrt{5}}}{\pi }\right)  \simeq 0.248137 .
\end{align}
This is a highly non-trivial evidence for the conjecture \eqref{SUSY enhancement}. 

\medskip 
\noindent
{\bf Evidence 2: $\mathcal{N}=4$ stress-energy tensor multiplet}

As another evidence, we construct $\mathcal{N}=4$ stress-energy tensor multiplet semi-classically. 

{\it Decomposition of $\mathcal{N}=4$ stress-energy tensor multiplet into $\mathcal{N}=2$ multiplets}:~The $\mathcal{N}=4$ supercharges are decomposed as
\begin{align}
\{ Q \}= \{ Q_{\mathcal{N}=2} \} \cup  \{ Q_{\rm IR} \},
\end{align}
where $Q_{\mathcal{N}=2}$ are the $\mathcal{N}=2$ supercharges manifest in the UV Lagrangian, 
while $Q_{\rm IR}$ denote the supercharges  emergent in the IR. 
$Q_{\mathcal{N}=2}$s are charged under $\UU(1)_R$ symmetry while $Q_{\rm IR}$s are charged under $\UU(1)_{\rm top}$ symmetry. In terms of multiplets of the $\mathcal{N}=2$ superconformal subalgebra (containing $Q_{\mathcal{N}=2}$ as supercharges), the $\mathcal{N}=4$ stress-energy tensor multiplet consists of
\begin{align}  
\begin{split}
&\boxed{\textrm{Conserved current multiplet, $A_2 \bar{A}_2[j=0]^{r=0}_{\Delta=1}$, for } \UU(1)_{\rm top} } 
\\
&\xrightarrow[\text{}]{ Q_{\rm IR}  } \boxed{\textrm{Two }A_1 \bar{A}_1 [j=\tfrac{1}{2}]^{r=0}_{\Delta = 3/2} \textrm{with $\UU(1)_{\rm top}$ charge $\pm 1$}} 
\\
&\xrightarrow[\text{}]{ Q_{\rm IR} }  \boxed{\textrm{Stress-energy tensor multiplet}, A_1\bar{A}_1 [j=1]^{r=0}_{\Delta=2}}.
\label{Table : N=4 multiplet}
\end{split}
\end{align}
Here $A_1 \bar{A}_1[j\geq \frac{1}2]^{r=0}_{\Delta = j +1}$ is a short-multiplet whose bottom component is a conformal primary with spin $j$, $\UU(1)$ R-charge $r$ and conformal dimension $\Delta=j+1$. We here follow the notation in \cite{Cordova:2016emh} with rescaling $j_{\rm our} = \frac{1}2 j_{\rm their}, r_{\rm our} = \frac{1}{2} r_{\rm their}$.  Both of the conserved current multiplet and the stress-energy tensor multiplet are expected to exist at the IR fix point assuming no dynamical breaking of the $\mathcal{N}=2$ supersymmetry and $\UU(1)_{\rm top}$ symmetry.  We only need to show the existence of an $A_1 \bar{A}_1 [j=\frac{1}2]^{r=0}_{\Delta = 3/2}$ multiplet at the IR fixed point in order  to show the emergent  $\mathcal{N}=4$ supersymmetry. 

{\it Semiclassical analysis on local operator spectrum}:~The local operator spectrum at semiclassical limit is summarized in Table~\ref{Table : local operator spectrum}.
\begin {table}[htbp]
\caption {Semiclassical local operators of the theory $\mathcal{T}_{k=-3/2,Q=1}$. For non-zero $q$, we only list operators which correspond to excited states on BPS monopole configuration and skip their conjugates.  
	We use the F-maximization result  in \eqref{IR R-charge} for the IR $\UU(1)_R$ charge. }
\begin{center}
	\begin{tabular}{| c| c | c | c | c|}
		\hline
		&  $\UU(1)_{\rm gauge}$ & $\UU(1)_{R}$ & $\UU(1)_{\rm top} $ & $(j,j_3)$  \\ \hline
		$a_{jm}^\dagger (j\in \frac{|q|}2 + \mathbb{Z}_{\geq 0})$&  $+1$ &  $\frac{1}3$ & 0 & $(j,m)$  \\ \hline
		$b^\dagger_{jm} (j\in \frac{|q|}2 + \mathbb{Z}_{\geq 0})$ &  $-1$ &  $-\frac{1}3$ & 0 & $(j,m)$  \\ \hline
		$\hat{a}^\dagger_{jm} (j\in \frac{|q|+1}2 + \mathbb{Z}_{\geq 0})$ &  $+1$ &  $-\frac{2}3$ & 0 & $(j,m)$  \\ \hline
		$\hat{b}^{\dagger}_{jm} (j\in \frac{|q|+1}2+ \mathbb{Z}_{\geq 0}) $ &  $-1$ &  $\frac{2}3$ & 0 & $(j,m)$  \\ \hline
		$\hat{c}^\dagger_{\frac{|q|-1}2,m}$ &  $+1 $ &  $-\frac{2}3$ & 0 & $(\frac{|q|-1}2,m)$  \\ \hline \hline
		$|q \rangle$ &  $ -\frac{|q|}2 - \frac{3q}2 $ &  $ \frac{|q|}3 $ & $q$ & (0,0)  \\ \hline
	\end{tabular} 
\end{center} \label{Table : local operator spectrum}
\end{table}
In the table, $|q\rangle$ denotes a 1/2 BPS holomorphic bare monopole operator with flux $q\in \mathbb{Z}$. Through a radial quantization, local operators of the 3d theory are mapped to  states on $S^2$. The bare monopole operator corresponds to a 1/2 BPS semi-classical configuration with 
\begin{align}
\int_{S^2} dA_{\UU(1)_{\rm gauge}} = 2\pi q.
\label{bare_monopole}
\end{align}
Upon the semi-classical configuration, we can excite bosonic and fermionic oscillators, $(a^\dagger, b^{\dagger})$ and $(\hat{a}^\dagger,\hat{b}^{\dagger},\hat{c}^\dagger)$, which come from following harmonic expansion (see e.g.\  \cite{Chester:2017vdh})
\begin{align}
&\phi =  \sum_{j \geq |q|} \sum_{m=-j}^j \bigg{(} a_{jm}^\dagger Y^*_{qjm} + b_{jm}Y_{qjm} \bigg{)},
\nonumber\\
&\psi =  \sum_{j \geq |q|+\frac{1}2} \sum_{m=-j}^j \bigg{(} \hat{a}_{jm}^\dagger A_{qjm}(\Omega_2) + \hat{b}_{jm}B_{qjm}\bigg{)} 
\nonumber\\
&\qquad + \sum_{m=\frac{1}2 - |q|}^{|q|-\frac{1}2} \hat{c}_{|q|-\frac{1}2,m} C_{q, |q|-\frac{1}2,m},
\end{align}
where the functions $Y$ are scalar monopole harmonics while $A,B$ and $C$ are spinor monopole harmonics under a proper normalization. In particular, the function $C$ corresponds to zero-modes of the Dirac operator on $S^2$ coupled to the monopole background \eqref{bare_monopole}.   
The bare monopole state is annihilated by all the annihilation operators
\begin{align}
\big{(}a_{jm}, b_{jm}, \hat{a}_{jm}, \hat{b}_{jm}, \hat{c}_{|q|-\frac{1}2, m}  \big{)} \cdot \big{|}q\big{\rangle}=0.
\end{align}
The bare monopole has  R-charge
$ \frac{|q|}3$ which is due to the zero-point shift from $|q|$ fermionic zero-modes $C$. The $\UU(1)$ gauge charge for the bare monopole come from two contributions, $- \frac{|q|}2$ from zero-modes and $- \frac{3}2 q$ from the classical Chern-Simons term with level $-\frac{3}2$. 

First, note that all gauge invariant operators have integer-valued $\UU(1)$ R-charges. The $\UU(1)$ R-symmetry is a subgroup of the non-Abelian  $\SO(4)$ R-symmetry  and its charge should be properly quantized (half-integers). 

Second, there is no gauge-invariant 1/2 BPS CPO in the semi-classical analysis. Bare monopole operator $|q\neq 0\rangle$ is not gauge-invariant and it needs to be dressed by matter fields by acting creation operators. Creation operators, except $\hat{c}^\dagger$ when $|q|=1$, have non-zero spin in the presence of monopole background and the excited gauge-invariant operators are not CPOs.  For $|q|=1$, the gauge charge of bare monopole can not be cancelled solely by $\hat{c}^\dagger$ and the excited gauge-invariant  operators are not CPOs.

{\it Superconformal index analysis}:~The 3d superconformal index \cite{Bhattacharya:2008zy,Dimofte:2011py} is defined as
\begin{align}
\mathcal{I}(u;x):=\textrm{Tr}\, (-1)^{r} x^{\frac{r}2 +j_3} u^{F}. 
\label{Index def}
\end{align}
Here  $u$ is the fugacity for the topological $\UU(1)_{\rm top}$ symmetry. 
The trace is taken over local operators, or states on $S^2$, of the theory. Contributions of most local operators are cancelled by its $Q$-transformed fermionic operator and only 1/4 BPS operators saturating the bound %
\begin{align}
\Delta \ge r+ j_3.
\end{align}  
could contribute to the index.
The index for 3d $\mathcal{N}=2$ theories can be computed using supersymmetric localization techniques \cite{Kim:2009wb,Dimofte:2011py}. For the theory $\mathcal{T}_{k=-3/2,Q=1}$, the index contains a power $x^{3/2}$:
\begin{align}
\begin{split}
\mathcal{I}^{\mathcal{T}_{-3/2,1}}(u;x) &= \sum_{ e  \in \mathbb{Z}}\mathcal{I}_{\Delta} \big{(}e, e;x \big{)} u^e,
\\
& = 1-x-\left(u+\frac{1}{u}\right) x^{\frac{3}{2}}  -2 x^2  +\dots , \label{SCI}
\end{split}
\end{align}
where the $\mathcal{I}_{\Delta}(m,e;x)$ is so-called tetrahedron index  \cite{Dimofte:2011py}
\begin{align}
\begin{split}
&\sum_{e\in \mathbb{Z}}\mathcal{I}_{\Delta}(m,e;x)u^e = \prod_{r=0}^{\infty} \frac{1-x^{r-\frac{m}{2}+1}u^{-1}}{1- x^{r-\frac{m}{2}} u}. 
\label{tetrahdron index}
\end{split}
\end{align}
Among all 3d $\mathcal{N}=2$ superconformal multiplets classified in \cite{Cordova:2016emh}, only the following two types of multiplets contributes a term $-x^{3/2}$ to the index
\begin{align}
A_1 \bar{A}_1 \left[j=\tfrac{1}{2}\right]^{r=0}_{\Delta = \frac{3}{2}} \textrm{ and } \, L \bar{B}_1\left[j=0 \right]^{r=3}_{\Delta = 3}.
\end{align}
The bottom component of $L\bar{B}_1$ corresponds to a CPO with R-charge $r$. The CPO contributes $(-1)^r x^{r/2}$ to the index. In  the descendant $Q_{\mathcal{N}=2}\cdot(A_1 \bar{A}_1[j=\frac{1}2]^{r=0}_{\Delta = 3/2})$, there is an operator with $\Delta = 2, r=1$ and $j=1$ which contributes $ (-x^{3/2})$ to the index.  
As seen in the above semi-classical analysis, there seems to be no CPO in the IR fixed point.  This means the term $-\left(u+\frac{1}{u}\right) x^{3/2}$ in the index should come from two $A_1 \bar{A}_1 [j=\frac{1}{2}]^{r=0}_{\Delta = 3/2}$  multiplets with $\UU(1)_{\rm top}$ charge $\pm 1$. 
This is compatible with the existence of the $N=4$ stress-energy tensor multiplet \eqref{Table : N=4 multiplet} in the IR SCFT. 

\medskip\noindent
{\bf Evidence 3: a duality between $(\mathcal{T}_{-3/2,1})^{\otimes 2}$ and  $T[\SU(2)]/\SU(2)^{\rm diag}_3$}

The third evidence for our conjecture comes from a duality derived from the 3d--3d correspondence \cite{Terashima:2011qi,Dimofte:2011ju,Lee:2013ida,Cordova:2013cea},
where a twisted compactification of 6d $A_1$ (2,0) theory on a 3-manifold $M$ generates an associated 3d $\mathcal{N}=2$ SCFT. When the 3-manifold $M$ has a  torus boundary, the resulting SCFT depends not only on $M$ but also on the choice of primitive boundary cycle $A \in H_1 (\partial M, \mathbb{Z}) = \mathbb{Z}\oplus \mathbb{Z}$. We denote the  3d SCFT by
\begin{align}
T[M;A]\;.
\end{align}
The theory have $U(1)_A$ symmetry associated the chosen boundary 1-cycle $A$. For a given 3-manifold,  there could be several topological representations and they give different UV descriptions which flow to the same IR fixed point. As a concrete example  relevent to our purpose,
let us consider a 3-manifold called  `figure-eight knot complement'
\begin{align}
M=(\textrm{figure-eight knot complement in $S^3$}).
\end{align}
The manifold has a torus boundary and  there is a canonical basis choice, $\mu$ (merdian) and $\lambda$ (longitude), for $H_1 (\partial M, \mathbb{Z})$
\begin{align}
H_1 (\partial M, \mathbb{Z}) = \mathbb{Z}\oplus \mathbb{Z} =\langle \mu, \lambda \rangle\;.
\end{align}
There are two well-known representations of this 3-manifold. One is using an ideal triangulation with two tetrahedra and the corresponding UV description $T^{\rm DGG}[M]$ is given as follows \cite{Dimofte:2011ju}
\begin{align}
\begin{split}
&T^{DGG}[M, A =\mu]
\\
&=(\textrm{a $\UU(1)$ vector multiplet  with vanishing CS level coupled} 
\\
& \qquad  \textrm{with two chiral multiplets $\Phi_1$ and  $\Phi_2$ of charge $+1$})\;.
\end{split}
\end{align}
The Lagrangian of the theory is given as follows in terms of superfields
\begin{align}
\begin{split}
&\mathcal{L}_{T^{DGG}[M,A =\mu]} = \int d^4 \theta( \Phi^\dagger_1 e^{V-V_\mu} \Phi_1 + \Phi_2^\dagger e^V \Phi_2 )
\\
&\qquad \qquad  +\frac{1}{4\pi} \int d^2 \theta (3 \Sigma_{V_\mu} V - \frac{3}2 \Sigma_{V_\mu} V_\mu)+(c.c)\;.
\end{split}
\end{align}
Here $V$ is the dynamical vector multiplet superfield for the $U(1)$ gauge symmetry and $\Sigma_V$ is its dual linear multiplet $\Sigma_V = \overline{D}^\alpha D_{\alpha }V$. $V_\mu$ is the background vector multiplet coupled to the $U(1)_\mu$ flavor symmetry. Then, the theory $T^{DGG}[M,A]$ with $A = \lambda$ is simply obtained by gauging the $U(1)_\mu$ symmetry of the above theory \cite{Dimofte:2011ju}. 
\begin{align}
\begin{split}
&\mathcal{L}_{T^{DGG}[M,A =\lambda]} = \int d^4 \theta( \Phi^\dagger_1 e^{V-W} \Phi_1 + \Phi_2^\dagger e^V \Phi_2 )
\\
&  +\frac{1}{4\pi} \int d^2 \theta (3 \Sigma_{W} V - \frac{3}2 \Sigma_{W} W +2 \Sigma_{W} V_\lambda)+(c.c)\;.
\end{split}
\end{align}
Here we renamed $V_\mu$ to $W$ and the vector multiplet $W$ is now dynamical. The $U(1)_\lambda$ symmetry in the theory corresponds to the $U(1)_{\rm top}$ symmetry for the  $U(1)$ gauge symmetry. Finally by redefining   the dynamical superfield $W$ to $V-W$
\begin{align}
\begin{split}
&\mathcal{L}_{T^{DGG}[M,A =\lambda]} = \int d^4 \theta( \Phi^\dagger_1 e^{W} \Phi_1 + \Phi_2^\dagger e^V \Phi_2 )
\\
&  +\frac{1}{4\pi} \int d^2 \theta \left(\frac{3}2 \Sigma_{V} V - \frac{3}2 \Sigma_{W} W +2 (\Sigma_{V}-\Sigma_W) V_\lambda \right)+(c.c)\;,
\end{split}
\end{align}
we see that
\begin{align}
\begin{split}
T^{\rm DGG}[M,A=\lambda] &= \mathcal{T}_{k=\frac{3}2,Q=1}\otimes \mathcal{T}_{k=-\frac{3}2,Q=1}
\\
& = (\mathcal{T}_{k=-\frac{3}2,Q=1})^{\otimes 2}. \label{DGG-theory}
\end{split}
\end{align}
Here $\mathcal{T}_1 \otimes \mathcal{T}_2$ means a decoupled product of two theories $\mathcal{T}_1$ and $\mathcal{T}_2$. 
In the second line, we used the triality which we will explain later in \eqref{triality}.
The other representation of the 3-manifold is
\begin{align}
\begin{split}
&M =  \big{(}\textrm{once-punctured torus bundle over $S^1$}
\\
&\qquad  \textrm{with monodromy } ST^3= \left(
	\begin{array}{cc}
	0 & 1 \\
	-1 & -3 \\
	\end{array}
	\right) \in SL(2,\mathbb{Z}) \big{)}.
\end{split}
\end{align} 
The 1-cycle around puncture  corresponds to the longitude cycle, $\lambda$.  Base on the representation, an alternative UV description $T^{\rm TY}[M]$ is proposed in \cite{Terashima:2011qi}:
\begin{align}
\begin{split}
&T^{\rm TY}[M;A = \lambda]= \frac{T[\SU(2)]}{\SU(2)^{\rm diag}_{3}}
\\
&:=\big{(}\textrm{gauging $\SU(2)^{\rm diag}$ of   $T[\SU(2)]$ with CS level $3$}\big{)}. 
\end{split}
\label{TY-theory}
\end{align}
Since the two theories arise from the same 3-manifold,
we expect following IR duality between two UV descriptions
\begin{align}
  \big{(}T^{\rm DGG}[M;A = \lambda] \textrm{ in \eqref{DGG-theory}}\big{)} 
  = \big{(} T^{\rm TY}[M ;A = \lambda]  \textrm{ in \eqref{TY-theory}}\big{)} \label{Duality-TY-DGG}.
\end{align}
The duality was checked by superconformal index in Ref.~\cite{Gang:2013sqa}.
The $T[\SU(2)]$ theory  \cite{Gaiotto:2008ak} is a theory living on a $S$-duality wall  in 4d $\mathcal{N}=4$ $\mathfrak{su}(2)$ maximally supersymmetric Yang-Mills theory, and is a 3d $\mathcal{N}=4$ $\UU(1)_0$ gauge theory coupled to two hyper multiplets of charge $+1$.
The theory has  manifest $\UU(1)_{\rm top} \times \SU(2)_H$ flavor symmetry  which are argued to be enhanced to $(\SU(2)_{\rm C}\times \SU(2)_H)/\mathbb{Z}_2$  symmetry at the IR fixed point from the self-mirror-dual property of theory.   

In the description of $T^{\rm TY}[M]$
we further deform the IR fixed point by gauging its diagonal subgroup $\SU(2)^{\rm diag}$ with CS level $+3$ \footnote{Such a diagonal gauging can also be applied  to generalized $T[SU(N)]$ theories and the gravity duals of the resulting non-Lagrangian 3d $\mathcal{N}=4$ theories are studied in \cite{Assel:2018vtq}}. 
In the gauging, we introduce an $\mathcal{N}=4$ $\mathfrak{su}(2)$ vector multiplet and add the following superpotential interaction \cite{Gaiotto:2007qi},
\begin{align}
\begin{split}
\delta W  &= -\frac{3}{4\pi} \textrm{Tr}[\Phi^2]+\textrm{Tr} [(\mu_C+\mu_H) \Phi],
\\
& \rightsquigarrow \frac{\pi}3 \textrm{Tr}[(\mu_C+\mu_H)^2].
\end{split}
\label{W_mu2}
\end{align}
Here $\Phi$ is the adjoint chiral field in  $\mathcal{N}=4$ vector multiplet, and $\mu_H$ and $\mu_C$ are holomorphic moment maps associated to the $\SU(2)_H$ and $\SU(2)_C$ flavor symmetries in  the $T[\SU(2)]$ theory respectively. 
 In the second line, we integrated out the massive adjoint chiral multiplet. 
 
Such a gauging  generically breaks the $\mathcal{N}=4$ supersymmetry to $\mathcal{N}=3$.
This is reflected in the breaking of the Cartan part of the R-symmetry,
from $\UU(1)_V\times \UU(1)_A$ to $\UU(1)_V$. Here $\UU(1)_V$ is the diagonal $\UU(1)$-subgroup of $\UU(1)\times \UU(1) \subset \SU(2)\times \SU(2)=\SO(4)_R$ while $\UU(1)_A$ is the anti-diagonal subgroup.  The $\UU(1)_V$ is actually the $\UU(1)_R$-symmetry of the $\mathcal{N}=2$ subalgebra. $(\mu_H,\mu_C)$ have charges $(+1,+1)$ under $\UU(1)_V=\UU(1)_R$ while $(+1,-1)$  under $\UU(1)_A$. 
Thus, the superpotential \eqref{W_mu2}
indeed does seem to break the $\UU(1)_A$ symmetry.

However, in our case we can appeal to the following properties of the moment map of $T[\SU(2)]$ theory  \cite{Gaiotto:2008ak}
\begin{align}
\textrm{Tr}[\mu_H^2] = \textrm{Tr}[\mu^2_C] =0,
\end{align}
so that the superpotential simplifies as
\begin{align}
\delta W  &= \frac{2\pi}{3} \textrm{Tr}[\mu_C\mu_H],
\label{W_mu_mu}
\end{align}
and hence enjoys the $\UU(1)_V\times \UU(1)_A$ symmetry. Actually the superpotential deformation  preserves full $SO(4)_R$ symmetry and $\mathcal{N}=4$ supersymmetry \footnote{Note that the superpotential deformation share the same group theoretical structure under the $\mathcal{N}=4$ superconformal algebra with the usual $\mathcal{N}=4$ supersymmetry preserving superpotential term $\textrm{Tr}[\phi_1 \Phi \phi_2]$.  Here $\Phi$ is an adjoint chiral multiplet in $\mathcal{N}=4$ vector multiplet and two chiral multiplets  $(\phi_1,\phi_2)$ in $(\textrm{fundamental} , \textrm{anti-fundamental})$ form a fundamental  hypermultiplet. In the comparision, $\Phi$ corresponds to $\mu_C$ and $\phi_1 \phi_2$ corresponds to $\mu_H$.}.

We have established that
the  two theories in  \eqref{Duality-TY-DGG} make manifest different amounts of supersymmetry
\begin{align}
\begin{split}
&\textrm{$T^{\rm DGG}[M] $ has manifest $\mathcal{N}=2$ supersymmetry},
\\
&\textrm{$T^{\rm TY}[M] $ has manifest $\mathcal{N}=4$ supersymmetry}.
\end{split}
\end{align}
Under a mild assumption that the manifest supersymmetries are not broken along the RG flow, 
the  duality \eqref{Duality-TY-DGG} implies that 
\begin{align}
\begin{split}
&\textrm{both theories have $\mathcal{N}=4$ supersymmetry in the IR},
\end{split}
\end{align}
namely that the theory $(\mathcal{T}_{k=-\frac{3}2,Q=1})^{\otimes 2}$, and hence $\mathcal{T}_{k=-\frac{3}2,Q=1}$ in itself \footnote{Suppose that the product SCFT $\mathcal{T}\otimes \mathcal{T}$, where $\mathcal{T}$ is an $\mathcal{N}\geq 2$ SCFT, has $\mathcal{N}\geq 3$ SUSY. $\mathcal{N}=3$ stress-energy tensor multiplet contains  an $\mathcal{N}=2$ $A_1 \bar{A}_1 [j=\frac{1}{2}]^{r=0}_{\Delta=3/2}$ multiplet. From the unitary bound, $\Delta \geq j+1-\frac{1}2(\delta_{j,0}+\delta_{j,1/2})$ for non-identity operator, we see that  the bottom component can not be constructed from a product of two non-identity operators in $\mathcal{T}$ unless $\mathcal{T}$ is a free-chiral theory. This implies that $\mathcal{T}$ itself should have the  multiplet and thus has $\mathcal{N} \geq 3$ SUSY. We can then use the superconformal index computation in  \ref{SCI} 
and appeal to sufficient condition for the supersymmetry enhancement criterion in Ref.~\cite{Evtikhiev:2017heo} and check that the $\mathcal{T}_{k=-3/2,Q=1}$ has $\mathcal{N}=4$ SUSY.}, 
has emergent $\mathcal{N}=4$ SUSY in the IR. This is a very strong evidence for the conjecture in  \eqref{SUSY enhancement}.

\medskip \noindent
{\bf More examples with $\mathcal{N}=4$ SUSY enhancement}

The 3d theory considered here allows infinitely many dual gauge theory descriptions (modulo topological sectors)  which are also expected to have enhanced SUSY in the IR. For example, following gauge theories are all equivalent as a 3d field theory on $\mathbb{R}^3$, where the  topological sector is invisible. 
\begin{align}
\begin{split}
&\mathcal{T}_{k=-\frac{3}2 p^2, Q=p} \quad \textrm{with $p\in \mathbb{Z}_{\neq 0}$},
\\
&\mathcal{T}_{k=\frac{3}2 p^2, Q=p}\quad \textrm{with $p\in \mathbb{Z}_{\neq 0}$},
\\
&\mathcal{T}_{k=0, Q=p} \quad \textrm{with $p\in 2\mathbb{Z}_{\neq 0}$}.
\label{triality}
\end{split}
\end{align}
The dualities among the three classes of gauge theories for $p=1$ above are studied in \cite{Gaiotto:2018yjh}.
The theories for general $p$ can be obtained  by rescaling the $\UU(1)$ gauge field $A_\mu$   by  $p A_{\mu}$. The rescaling does not affect the spectrum of local operators modulo rescaling of (electric charge $m$, magnetic charge $e$)  to $(\frac{1}p m, p e)$ due to the change of Dirac quantization. The rescaling does not break any (0-form) symmetry including supersymmetry but  breaks some 1-form symmetries \cite{Gaiotto:2014kfa}. 
 
 Further, the theory has a realization in the context of 3d--3d correspondence, where the 3-manifold in question is a closed hyperbolic 3-manifold whose hyperbolic volume is $\textrm{Im}(\textrm{Li}_2 (e^{\frac{i \pi}3})) =1.01494$ \cite{Gang:2017lsr}. From infinitely many  Dehn surgery representations of the 3-manifold,  we have  infinitely many dual  descriptions of the abelian gauge theory \cite{Gang:2018wek}. 

\medskip
We would like to thank Victor Mikhaylov for collaboration in a related exciting project \cite{to-appear}, from which 
this project emerged as an excitement enhancement. We also thank Kimyeong Lee, Piljin Yi, Sangmin Lee, Sungjay Lee, Jaemo Park, Seok Kim, Jaewon Song, Prarit Agarwal, Kazuya Yonekura and Zohar Komargodski for useful comments.
The work of DG was supported by Samsung Science and
Technology Foundation under Project Number SSTBA1402-08. 
The work of MY was supported by JSPS Grant No. 17KK0087.

\bibliographystyle{ytphys}
\bibliography{ref}

\end{document}